\documentclass[smallabstract,smallcaptions]{dccpaper}

\usepackage{graphicx}
\usepackage{amsmath}
\usepackage{amssymb}
\usepackage{booktabs}
\usepackage[hyphens]{url}
\usepackage{color}
\usepackage{placeins}
\newcommand{\rate}[2]{r_{#1}(#2)}
\newcommand{\dist}[2]{d_{#1}(#2)}
\newcommand{\model}{\hat{p}}

\begin{document}

\title
{\large
\textbf{Co-Adaptive Entropy Modeling and\\
Rate--Distortion-Optimal Bit Allocation\\
for Perceptual Audio Coding}
}

\Support{Preprint of a manuscript submitted to the 2027 Data Compression Conference (DCC 2027). This work has been submitted to the IEEE for possible publication. Copyright may be transferred without notice, after which this version may no longer be accessible.}
\Notes{$^{\ddag}$Soohyun Kim and Jiwon Jeong contributed equally to this work.}

\author{%
Soohyun Kim$^{\ast\ddag}$, Jiwon Jeong$^{\dag\ddag}$, and Marina Bosi$^{\ast}$\\[0.5em]
{\small
\begin{tabular}{@{}c@{\hspace{0.35in}}c@{}}
$^{\ast}$Center for Computer Research & $^{\dag}$Department of Electrical Engineering \\
in Music and Acoustics (CCRMA) & Stanford University \\
Stanford University & Stanford, CA 94305, USA \\
Stanford, CA 94305, USA & \url{jeongjw@stanford.edu} \\
\url{{soohyun,mab}@ccrma.stanford.edu} & \\
\end{tabular}
}
}

\maketitle
\thispagestyle{empty}
\suppressfloats[t]

\begin{abstract}
We present a perceptual audio coder in which the entropy-coding
model and the bit allocation co-adapt across blocks without model side
information. Mantissas are coded by a hybrid of bit packing and arithmetic
coding with per-band probability models learned online. Encoder and decoder
accumulate identical histograms from decoded blocks and advance the same model
state. The model enters bit allocation directly: per-band rate--distortion frontiers combine measured
perceptual distortion with entropy-coded rate estimates under the current
model, and a dynamic program attains the exact minimum of total
objective perceptual distortion for each block's budget under the model's rate
estimates, with reservoir carry-over. Under a strict per-block rate,
entropy-coding-aware bit allocation lowers this distortion (band
noise-to-mask ratio in linear scale) by 15--24\,\% relative to the same solver
without it, and co-adaptation removes a further 3--4\,\%
(up to 15\,\% on oboe) that a backward-adaptive coder alone cannot realize in
one pass; a MUSHRA test at 80 kbps per channel confirms the ranking.
\end{abstract}

\Section{Introduction and Background}

Perceptual coders shape quantization noise below the masked
threshold of a psychoacoustic model and typically allocate a fixed or locally-variable bit budget across
frequency bands
\cite{johnston1988transform,brandenburg1994iso,bosi2003book,herre2025perceptual}.
To improve the performance of such coders, we combine two research lines.
The first is \emph{allocation}: perceptually motivated bit allocation dates
to \cite{johnston1988transform,voran1997perception}---Voran's allocations use
a raw bits-times-lines rate model with no entropy coding---and discrete
rate--distortion allocation with general rate models is classical
\cite{shoham1988efficient,riskin1991optimal}.
Closest to our solver layer, Aggarwal \emph{et al.}\ \cite{aggarwal2006trellis}
select MPEG Advanced Audio Coding (AAC) \cite{bosi1997aac} scale factors
and Huffman codebooks by a trellis search minimizing average or maximum
noise-to-mask ratio (NMR), exact for each Lagrange multiplier, with rates
taken from the \emph{static, standardized} Huffman tables. Oger \emph{et al.}\ set the quantizer step from a
generalized-Gaussian spectrum model fitted at the encoder and code the
coefficients with an adaptive stack-run arithmetic coder, but the coder's
adaptive state does not price the allocation \cite{oger2007transform}.

The second line is \emph{adaptive entropy coding}: Context-based Adaptive Binary Arithmetic Coding (CABAC) \cite{marpe2003cabac},
MPEG Unified Speech and Audio Coding (USAC)'s context-adaptive arithmetic coder
(CAAC) \cite{fuchs2011efficient},
which switches among
fixed trained context tables, and time--frequency context arithmetic coding
for G.719 \cite{wang2013context} adapt per symbol from decoded context; in
the audio coders among them, the adaptation state is not used to price
allocation candidates. Malvar's run-length/Golomb--Rice coder and
lapped-transform audio coder likewise adapt from previously decoded data
without side information \cite{malvar2006adaptive,malvar2007lossless}, within
the entropy layer only. Conversely, AC-3's ``hybrid backward/forward
adaptive'' allocation runs the same core allocation in encoder and decoder
from the decoded spectral envelope \cite{davidson1994parametric},
but with no entropy coding in the loop. To our knowledge, no perceptual audio
coder closes the loop between a backward-adaptive entropy model and a
discrete rate--distortion-optimal bit-allocation stage: if the entropy model
is \emph{learned online from decoded data}, the bit-allocation stage can be
made aware of the \emph{current} coded cost of every candidate quantization
choice---at zero side-information cost.

Our contribution is this closed loop: a decoder-synchronized entropy model is
coupled to entropy-coding-aware perceptual bit allocation, so that the two
co-adapt across blocks without transmitting model parameters. Concretely:
(i) \textbf{Entropy-coding-aware allocation.} Per-band rate--distortion
frontiers are built from \emph{measured} quantization distortion (weighted by
the masked threshold) and \emph{entropy-coded} rate estimates under the
current arithmetic model.
(ii) \textbf{Exact and approximate solvers.} A trellis dynamic program (DP)
attains the exact minimum of total objective perceptual distortion for each
block's budget under the model's rate estimates; we compare it to a
marginal-analysis solver in the style of the generalized
Breiman--Friedman--Olshen--Stone (BFOS) algorithm
\cite{riskin1991optimal} and a Lagrangian solver
\cite{shoham1988efficient}, exposing a quality--complexity
trade-off. (iii) \textbf{Causal co-adaptation.} Encoder and
decoder advance a shared per-(bit precision, band) histogram model from
decoded mantissas; a control with the coder adapting but the allocation costs
frozen isolates the allocation feedback. (iv) \textbf{Validation.} Objective
distortion and CPU time on six standard excerpts at 64 and 80 kbps/channel,
and a listening test in the style of Multiple Stimuli with Hidden Reference
and Anchor (MUSHRA) \cite{bs1534} at 80 kbps/channel. Under a strict per-block
rate, entropy-coding-aware bit allocation lowers total objective perceptual
distortion by 15--24\,\% relative to the same solver without it, and
co-adaptation removes a further 3--4\,\% (up to 15\,\% on oboe)
that a backward-adaptive coder alone cannot realize in one pass; with an unbounded
reservoir the increments shrink to 1--3\,\% and 0.3--0.5\,\%, the complete
system reaches 57--63\,\% of conventional water-filling, and the listening
test places the exact-DP conditions 22--25 MUSHRA points above water-filling,
with a further 3.4 points for co-adaptation.

\Section{Codec Overview}\label{sec:overview}

The coder processes pulse-code-modulated (PCM) audio; in the current implementation, the two channels of a 16-bit stereo signal sampled at 44.1 kHz are
coded independently (no joint-stereo tools), and all rates are quoted per
channel. Each channel is analyzed with a sine-windowed modified discrete cosine
transform (MDCT) \cite{princen1987subband} of $N{=}1024$ frequency lines
(2048-sample blocks, 50\% overlap). Lines are grouped into $K{=}50$
scale-factor bands with equal width on the equivalent-rectangular-bandwidth
(ERB) rate scale \cite{glasberg1990derivation};
band $k$
holds $L_k$ lines. A block floating-point quantizer with 4-bit scale factors
and per-band mantissa precision $b_k\in\{0\}\cup\{2,\dots,16\}$ quantizes each
band. A side-chain fast Fourier transform (FFT)-based psychoacoustic model
(tonal maskers, threshold in quiet \cite{bosi2003book}) yields per-band
signal-to-mask ratios $\mathrm{SMR}_k$; the masking curve rises toward the
masker at 27\,dB/Bark and falls above it at $27-0.367(L{-}40)$\,dB/Bark, where
$L$ is the masker level in dB; the tested implementation leaves $L{-}40$
unclamped. Per block and
channel, the encoder builds per-band rate--distortion frontiers, solves the
allocation, and writes the bitstream with per-band arithmetic-coding
(AC)/bit-packing selection; both sides then update the shared entropy model,
and unspent bits enter a reservoir.

\Section{Hybrid Entropy Coding Layer}\label{sec:entropy}

Arithmetic coding \cite{witten1987arithmetic,odland2024arithmetic} encodes both scale factors and mantissas. Each AC payload has an 8-bit header giving its length in bits, which keeps the overhead low, followed by the coded symbols; payloads longer than 255 bits fall back to bit packing. For the scale factors of a block and the mantissas of each band, the encoder selects the shorter of bit packing and arithmetic coding, signaling each choice with one flag bit.

Scale factors are encoded as differences between consecutive bands under a static, empirically derived probability model; neighboring bands tend to have similar scale factors, giving the differences lower entropy.

Each (bit precision, frequency band) pair has an initial mantissa probability model: precision sets the alphabet size, and distributions vary by band. Each model uses a maximum-likelihood double-geometric fit, with the same geometric distribution for positive and negative mantissas. This model is updated in the co-adaptive mode, as described in the \emph{Causal Co-Adaptation of Model and Allocation} section. Because lower frequency bands have fewer mantissas per band, plain bit packing tends to beat arithmetic coding with its 8-bit header there, so bit packing is used for bands 0 to 34 and the hybrid option is enabled from band 35, an empirically chosen threshold.

\Section{Entropy-Coding-Aware Rate--Distortion Frontiers}\label{sec:alloc}

Fix a block $t$ and channel. Let $\mathbf{x}_k\in\mathbb{R}^{L_k}$ denote the
vector of MDCT coefficients on the $L_k$ spectral lines of band $k$, and let
$b_k$ denote the number of mantissa bits assigned to band $k$, to be selected
by the bit-allocation stage from the candidate set $\mathcal{B}_k=\{0\}\cup\{2,\dots,15\}$
in the arithmetic-coded region and up to 16 elsewhere (a precision of 1 is invalid
for the midtread quantizer). For each candidate $b_k$, write
$\hat{\mathbf{x}}_k(b_k)\in\mathbb{R}^{L_k}$ for the \emph{reconstruction} of
$\mathbf{x}_k$ by the codec's block floating-point quantizer/dequantizer with 4-bit overall block
and per-band scale factors, with $\hat{\mathbf{x}}_k(0)=\mathbf{0}$.

\SubSection{Objective measure of perceptual distortion}
We use a perceptual distortion metric that normalizes the squared reconstruction error of each candidate by the band's squared masking threshold. For band $k$, let $a_k=\max_j |x_{k,j}|$ denote the peak amplitude and $m_k = a_k\,10^{-\mathrm{SMR}_k/20}$ the mask amplitude. We define
\begin{equation}
\dist{k}{b_k} \;=\; \frac{1}{m_k^{2}}\,
\bigl\lVert \mathbf{x}_k - \hat{\mathbf{x}}_k(b_k)\bigr\rVert^2 .
\label{eq:dist}
\end{equation}
Silent bands have $\dist{k}{b_k}=0$. This \emph{objective perceptual distortion} is
the band's noise-to-mask ratio (NMR) in linear scale, summed over the $L_k$ lines
of band $k$. Note that this quantity is invariant to the overall block scale. Minimizing its sum over all bands $k$---the \emph{total objective
perceptual distortion}---is the allocation objective of the \emph{Allocation Solvers}
section.

\SubSection{Model-dependent rate}
The rate of a candidate is the smaller of plain bit packing and the
entropy-coded cost under the \emph{current} probability tables $\model^{(t)}$,
updated as described in \emph{Causal Co-Adaptation of Model and Allocation}:
\begin{equation}
\rate{k}{b_k} \;=\;
\begin{cases}
0, & b_k = 0,\\[2pt]
b_k\,L_k, & k < k_0,\\[2pt]
\min\!\Bigl(b_k\,L_k,\;
\bigl\lceil h + \textstyle\sum_{j} -\log_2
\model^{(t)}_{k,b_k}\!\left(q_{k,j}(b_k)\right) \bigr\rceil\Bigr), & k \ge k_0,
\end{cases}
\label{eq:rate}
\end{equation}
where $q_{k,j}(b_k)$ are the integer mantissa symbols produced by the
quantizer (whose dequantization yields $\hat{\mathbf{x}}_k(b_k)$), $h{=}8$ is
the per-band payload length header, and $k_0{=}35$ is the first
arithmetic-coded band. The cross-entropy estimate in \eqref{eq:rate} avoids
trial encoding but omits the coder's termination bits and the 255-bit payload
cap, so it is a surrogate for the written cost: it typically tracks the
written payload to within a few bits per block, and the reservoir absorbs the
residual (see \emph{Reservoir coupling} below). Every
candidate precision thus yields one rate--distortion point
$(\rate{k}{b_k},\dist{k}{b_k})$ for band $k$, and each band's point set is
Pareto-pruned: a candidate is discarded whenever another candidate of the same
band has no higher rate and strictly lower distortion, leaving a compact
per-band frontier $\mathcal{F}_k$. Pruning does not change the optimum of
the allocation problem \eqref{eq:tdproblem} introduced in the next section, because a dominated
choice can always be replaced by the candidate that dominates it without
increasing rate or distortion. The frontiers are Pareto-pruned but not
necessarily convex-pruned, which is what separates the exact solver from the
greedy and Lagrangian ones.

\Section{Allocation Solvers}\label{sec:solvers}

The per-channel mantissa budget available to block $t$ is
\begin{equation}
B = \max\!\left(0,\left\lfloor R_0-S+v^{(t)}/C\right\rfloor\right),
\label{eq:budget}
\end{equation}
where $R_0$ is the target bits per channel and block, $S=420$ reserves scale
factors, allocations and flags, and $v^{(t)}$ is shared by the $C$ channels.

\SubSection{Exact dynamic program for total objective perceptual distortion}
Consistent with \cite{aggarwal2006trellis}, minimizing the \emph{total}
(average) objective perceptual distortion correlates better with subjective quality
than the classic minimax-NMR objective; we therefore seek
\begin{equation}
\min_{(b_1,\dots,b_K)}\;\sum_{k=1}^{K} \dist{k}{b_k}
\qquad \text{s.t.} \quad \sum_{k=1}^{K} \rate{k}{b_k} \le B .
\label{eq:tdproblem}
\end{equation}
Define $V_k(u)$ as the minimum accumulated objective perceptual distortion over bands
$1,\dots,k$ using exactly $u$ bits, for integers $0\le u\le B$. With no bands,
only zero bits is reachable: $V_0(0)=0$ and $V_0(u{>}0)=\infty$. The recurrence
and terminal minimum are
\begin{equation}
\begin{aligned}
V_k(u) &= \min_{b_k\in\mathcal{B}_k,\; \rate{k}{b_k}\le u}
\Bigl[ V_{k-1}\bigl(u-\rate{k}{b_k}\bigr) + \dist{k}{b_k} \Bigr],\\
D^\star(B) &= \min_{0\le u\le B} V_K(u).
\end{aligned}
\label{eq:dp}
\end{equation}
Ties favor smaller $u$; backtracking recovers the allocation. Since rates are integers, the
trellis over $(k,u)$ states provides an exact solution to problem \eqref{eq:tdproblem}---no
convexity assumption on $\mathcal{F}_k$ is required. The guarantee holds for the
block's budget under the rate estimates \eqref{eq:rate}: written bits can
differ by the residual described above, and the block sequence is handled causally
through the reservoir, not jointly optimized. Given the mean frontier size $\bar F$ ($\le 16$), the complexity is
$O(KB\bar F)$ in time and $O(KB)$ in traceback memory.

\SubSection{Marginal analysis (BFOS-style) and Lagrangian relaxation}
The marginal-analysis solver, in the style of the generalized BFOS
algorithm \cite{riskin1991optimal}, starts each band at its
cheapest point and repeatedly takes the affordable step of largest
$-\Delta d/\Delta r$; unlike generalized BFOS it does not first take the lower
convex hull of each frontier, and with unequal, indivisible rate steps even
convex frontiers do not guarantee optimality, so it is a heuristic for these
discrete frontiers. The Lagrangian solver
\cite{shoham1988efficient} minimizes
$\sum_k \dist{k}{b_k}+\lambda \rate{k}{b_k}$ per band, bisects on $\lambda$
(30 iterations) to meet $B$, then greedily projects remaining slack onto the
steepest steps.
All solvers use the same frontier construction, so their comparisons isolate
the optimization method from the cost model (in complete runs the reservoirs,
and with co-adaptation the models, evolve separately).

\SubSection{Reservoir coupling}
After coding, $v^{(t+1)} = v^{(t)} + C\,R_0 - R_{\mathrm{used}}^{(t)}$, where
$R_{\mathrm{used}}^{(t)}$ is the coded payload for block $t$ across channels.
Entropy-coded rates let the solver use coding savings to reduce the current
block's distortion in one pass. There is no stop-at-mask condition; unused bits,
scale-factor savings and rate-estimation residuals enter the reservoir.
In our entropy-unaware water-filling control, coding savings instead affect
later blocks through the reservoir. With an unbounded reservoir, payload rates average 63.99 and
79.99 kbps/ch, and every stream ends with a non-negative reservoir.
File headers, 32-bit per-channel block-length fields, byte alignment and final
flush blocks add about 1.7--1.8 kbps/ch, excluded from nominal rates. The prototype does not constrain the instantaneous rate;
Table~\ref{tab:feedback} also evaluates bounded reservoir policies.

\Section{Causal Co-Adaptation of Model and Allocation}\label{sec:coadapt}

The model state is a table of per-(precision, band) symbol histograms
$\mathcal{H}^{(t)} = \{H^{(t)}_{b,k}\}$ where $b$ and $k$ denote precision and band, respectively; the coder starts from a fixed
double-geometric fit and the observed counts start empty. After block $t$ is
coded, \emph{both} encoder and decoder execute the identical update from the
decoded mantissas $q^{(t)}$:
\begin{equation}
H^{(t+1)}_{b,k} \;=\; H^{(t)}_{b,k} + \sum_{\text{ch}}
\mathrm{hist}\bigl(q^{(t)}_{k}\bigr)\cdot
\mathbf{1}\bigl[b_k^{(t)}{=}b\bigr],
\qquad k\ge k_0,\; 2\le b\le b_{\max},
\label{eq:update}
\end{equation}
applied on a window cadence: accumulated counts are merged into the model
every $T{=}10$ blocks, and a cell is (re)built only once its window count
reaches $\tau{=}20$ samples (smaller windows are discarded); precisions are
capped at $b_{\max}{=}13$ to keep the bands with large alphabets stable. The first merged
window replaces the parametric prior of its cell with the learned
nonparametric table (integer frequencies floored at one and rescaled to the
coder's precision), later windows accumulate without forgetting, and untouched
cells retain the prior. The frequency tables of the arithmetic coder
\emph{and} the rate terms \eqref{eq:rate} of the bit-allocation stage are rebuilt from
$\mathcal{H}^{(t+1)}$, so the next block's allocation is optimized against the
model the coder will actually use. Windowing ($T$), the count threshold
($\tau$), and the precision cap ($b_{\max}$) damp the feedback between
allocation and model; we observed no oscillatory behavior in any test item.

\SubSection{Synchronization invariant}
The update rule \eqref{eq:update} depends only on quantities the decoder
reproduces exactly: the transmitted allocations $b_k^{(t)}$ and the decoded mantissas.
By induction, if both sides start from the same preinstalled base model (the
co-adaptive mode is signaled in the stream header), then $\mathcal{H}^{(t)}$---and
hence every arithmetic table---is identical on both sides for all $t$, and
\emph{no model side information is ever transmitted}. This invariant is what
makes nonparametric per-$(b,k)$ tables practical at all: transmitting
$12\times15$ learned cells with up to $2^{13}$ symbols forward would dwarf the
payload.

\SubSection{Codec complexity}
Frontier construction quantizes and prices every candidate precision before
pruning. Lagrangian search costs $O(IK\bar F)$ for $I$ multiplier evaluations;
marginal analysis and the final greedy fill scan all bands per step, costing
$O(K^2\bar F)$. At 80 kbps/channel, DP takes about 20 times the Lagrangian
encode CPU time for about 0.5\,\% lower distortion (Table~\ref{tab:objective}).
Co-adaptation accumulates histograms each block and rebuilds coder and rate
tables only for changed cells every $T$ blocks; its average encoding overhead
over static AC is about 1\,\%. The decoder adds histogram accumulation and occasional table rebuilds.

\Section{Experimental Results}\label{sec:results}

\SubSection{Setup}
Six excerpts from the European Broadcasting Union (EBU) Sound Quality
Assessment Material (SQAM) disc \cite{ebu2008sqam} (oboe, glockenspiel, castanets, German male speech,
orchestra, soprano) were selected to evaluate the system at 64 and 80 kbps per channel.
Conditions are labeled \emph{solver (entropy mode)} as defined in
Table~\ref{tab:objective}; all share the MDCT filterbank, psychoacoustic model,
and quantizer, and the AC conditions also share the bitstream syntax, so only
the allocation solver and the entropy model differ. The initial entropy model
was fitted on SQAM material that includes four of the six excerpts
(glockenspiel, speech, castanets, oboe); all conditions share this model, and
we report these four as \emph{in-domain} and orchestra and soprano, which
were not used in the fit, as \emph{out-of-domain}. Encode and decode CPU times were measured on a single core.

\begin{figure}[!tb]
\centering
\includegraphics[width=\linewidth]{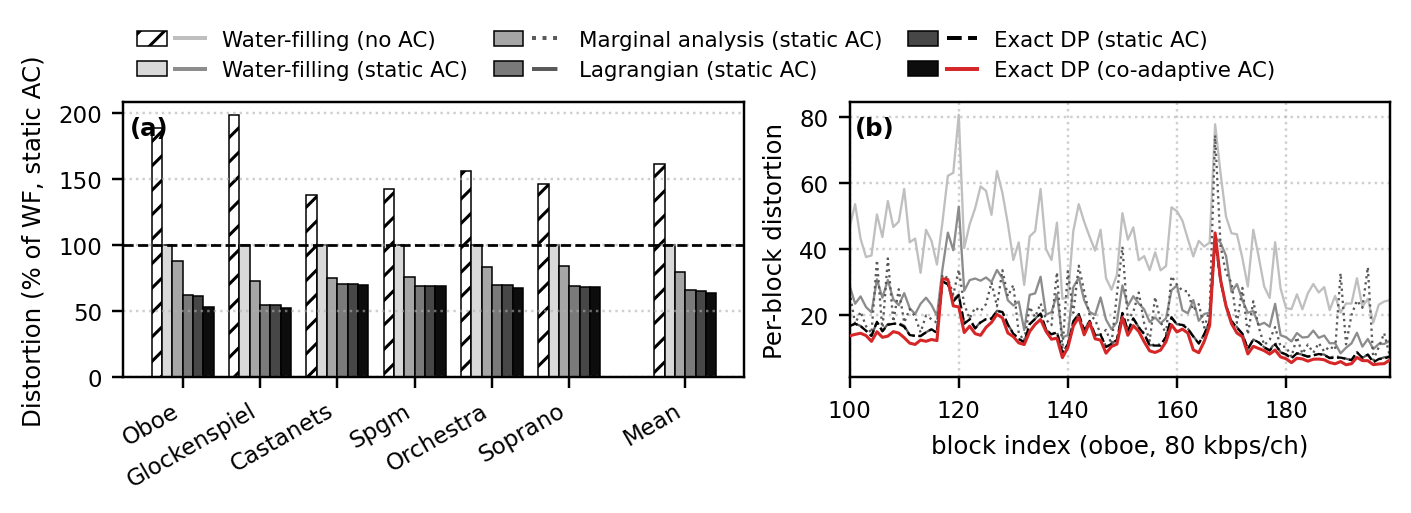}
\caption{\label{fig:objective}%
(a) Mean total objective perceptual distortion (both channels, unbounded
reservoir) per excerpt at 80 kbps/channel as a percentage of Water-filling
(static AC) (dashed line at 100). (b)
Per-block total objective perceptual distortion on the oboe excerpt (blocks
100--200; Lagrangian and Exact DP (static AC) overlap): once the co-adaptive
model has adapted, Exact DP (co-adaptive AC), solid red, tracks below Exact DP
(static AC), black dashed, in nearly every block.}
\end{figure}

\SubSection{Objective results}
Table~\ref{tab:objective} reports the objective of \eqref{eq:tdproblem}
relative to Water-filling (static AC), and Fig.~\ref{fig:objective} breaks
the 80 kbps/channel result down by excerpt and block. Without entropy coding,
distortion rises to 139\,\% (64 kbps/ch) and 162\,\% (80 kbps/ch) of
Water-filling (static AC). The exact DP on entropy-coding-aware frontiers
reaches 59--66\,\% of that baseline, the co-adaptive model removes a further 3--4\,\% for the
Lagrangian and the exact DP, and the Lagrangian solver stays within about
0.5\,\% of the exact DP at about $1/20$ of its encode CPU time under both
entropy modes.

\emph{Role of entropy-coding awareness.} The gap between water-filling and
the exact DP mixes two changes: allocating on measured, mask-weighted
distortion with an exact solver, and pricing each candidate at its
entropy-coded rate. A control that runs the same solver with every candidate
priced at its uncoded length $b_k L_k$ separates them. Under a strict
per-block rate (Table~\ref{tab:feedback}), entropy-coding awareness lowers distortion by 15\,\% (64
kbps/ch) and 24\,\% (80 kbps/ch) relative to that control, up to 39\,\% on
oboe, in- and out-of-domain alike; with a reservoir of one block or more, the
reservoir forwards the mispriced bits to later blocks and the increment
shrinks to 1--3\,\%. The remaining 33--40 points of the gap come from the
measured-distortion frontiers and the exact solver.

\emph{Role of co-adaptation.} Co-adaptation changes both the arithmetic
coder's tables and the rate terms of the bit-allocation stage; the coder-only
control of Table~\ref{tab:feedback} separates the two. Under a strict
per-block rate, co-adaptation retains the 3.3--3.6\,\% reduction (up to
15\,\% on oboe; 1.6--1.7\,\% out-of-domain); once the reservoir can carry
even one block's budget forward, it delivers most of the savings to later
blocks by itself and the allocation feedback adds a further 0.3--0.5\,\%. The feedback
thus spends the coder's savings in the block where they arise, so that the
co-adaptive gain is realized under every tested reservoir policy;
Fig.~\ref{fig:objective}(b) shows it block by block on oboe.

\begin{table}[!tb]
\begin{center}
\caption{\label{tab:objective}%
Mean total objective perceptual distortion over the six excerpts (both
channels, unbounded reservoir), as \% of Water-filling (static AC) per
excerpt, averaged over excerpts; lower is better. Static AC = fixed pretrained entropy model;
co-adaptive AC = model and allocation costs updated from decoded data; no AC
= plain bit packing with the same reservoir policy. Encode and decode CPU times are relative to
Water-filling (static AC) at 80 kbps/ch (single-threaded Python prototype).}
{\renewcommand{\baselinestretch}{1}\footnotesize
\begin{tabular}{lcccc}
\toprule
Condition & 64 kbps/ch & 80 kbps/ch & Enc.\ CPU & Dec.\ CPU \\
\midrule
Water-filling (no AC)             & 138.8 & 161.7 & 0.8$\times$ & 0.6$\times$ \\
Water-filling (static AC)         & 100.0 & 100.0 & 1$\times$ & 1$\times$ \\
Marginal analysis (static AC)     & 67.7  & 79.9  & 4.1$\times$ & 1$\times$ \\
Lagrangian (static AC)            & 59.5  & 66.0  & 4.1$\times$ & 1$\times$ \\
Exact DP (static AC)              & 59.3  & 65.7  & 88$\times$ & 1$\times$ \\
Lagrangian (co-adaptive AC)       & 57.7  & 63.7  & 4.2$\times$ & 1$\times$ \\
\textbf{Exact DP (co-adaptive AC)} & \textbf{57.5} & \textbf{63.5} & 88$\times$ & 1$\times$ \\
\bottomrule
\end{tabular}}
\end{center}
\end{table}

\begin{table}[!tb]
\begin{center}
\caption{\label{tab:feedback}%
Controlled increments for the exact DP under three reservoir policies. Control
and Proposed: total objective perceptual distortion (both channels) as \% of
Water-filling (static AC) under the same policy, mean over the six excerpts, at
64\,/\,80 kbps/ch. Gain: relative reduction from control to proposed per excerpt,
mean over the six excerpts. The coder-only control adapts the entropy coder
as a self-contained lossless layer, as prior backward-adaptive coders do
\cite{fuchs2011efficient,wang2013context,malvar2006adaptive,malvar2007lossless},
without updating allocation costs.}
{\renewcommand{\baselinestretch}{1}\footnotesize
\begin{tabular}{lccc}
\toprule
Reservoir & Control & Proposed & Gain (\%) \\
\midrule
\multicolumn{4}{l}{\emph{Entropy-coding awareness}: uncoded rates $\to$ entropy-coded rates} \\
None (strict per-block rate) & 58.0\,/\,62.7 & 49.5\,/\,48.1 & 15.0\,/\,23.7 \\
One block budget & 60.7\,/\,67.5 & 59.7\,/\,66.3 & 1.6\,/\,1.8 \\
Unbounded (as in Table~\ref{tab:objective}, Figs.~\ref{fig:objective}--\ref{fig:mushra}) & 60.1\,/\,66.8 & 59.3\,/\,65.7 & 1.4\,/\,1.7 \\
\midrule
\multicolumn{4}{l}{\emph{Co-adaptation}: coder only $\to$ co-adaptive} \\
None (strict per-block rate) & 49.5\,/\,48.1 & 48.2\,/\,46.7 & 3.3\,/\,3.6 \\
One block budget & 58.2\,/\,64.4 & 58.1\,/\,64.2 & 0.3\,/\,0.3 \\
Unbounded (as in Table~\ref{tab:objective}, Figs.~\ref{fig:objective}--\ref{fig:mushra}) & 57.7\,/\,63.6 & 57.5\,/\,63.5 & 0.4\,/\,0.3 \\
\bottomrule
\end{tabular}}
\end{center}
\end{table}

\begin{figure}[!tb]
\centering
\includegraphics[width=\linewidth]{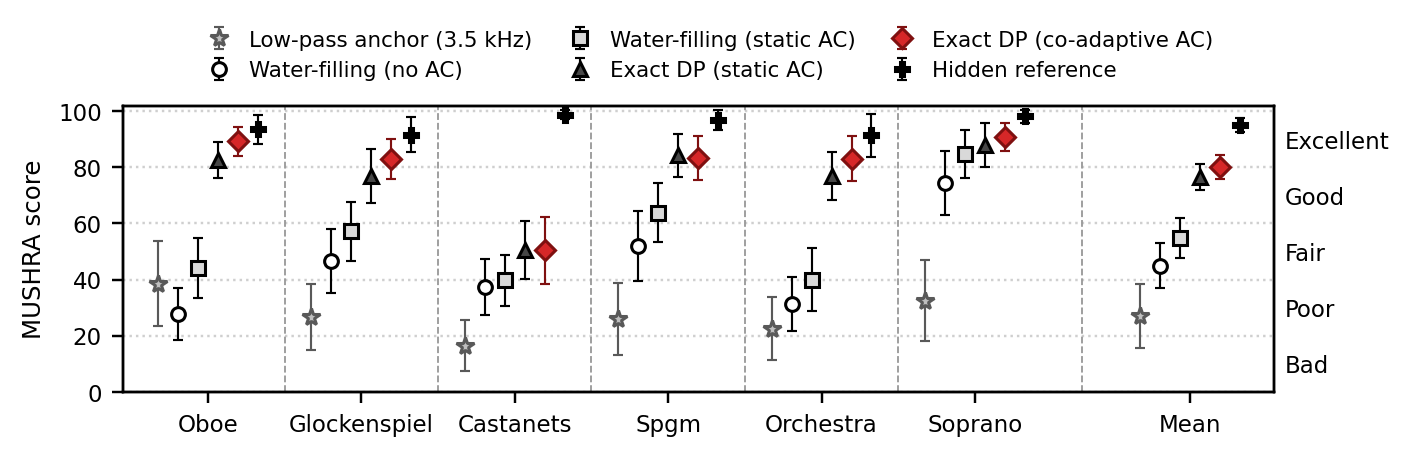}
\caption{\label{fig:mushra}%
Subjective listening test results: mean MUSHRA scores with 95\% confidence
intervals across 20 listeners, per excerpt and overall, at 80 kbps/channel
(AC conditions use an unbounded reservoir).}
\end{figure}

\SubSection{Subjective listening test results}
A MUSHRA test \cite{bs1534} with a hidden reference and a single 3.5 kHz
low-pass anchor (no 7 kHz mid anchor) compared Water-filling (no AC), Water-filling (static
AC), Exact DP (static AC), and Exact DP (co-adaptive AC) on the six excerpts
at 80 kbps/channel; the no-AC stimuli used an earlier configuration without a reservoir or strategy flags
and have about 3\,\% lower objective distortion than the no-AC condition in
Table~\ref{tab:objective}. All 20 listeners were retained; they used the same Sennheiser HD~600
headphones in a listening room after familiarization with standard coding
artifacts, with the hidden conditions randomized within each trial. Mean
scores were 44.9 for Water-filling (no AC), 55.0 for Water-filling (static
AC), 76.6 for Exact DP (static AC), and 80.0 for Exact DP (co-adaptive AC),
compared to 95.1 for the reference and 27.1 for the anchor
(Fig.~\ref{fig:mushra}). On paired listener means, co-adaptive minus static
DP is $+3.4$ points (95\,\% confidence interval 0.3--6.5, paired $t$-test
$p=.035$).

\Section{Conclusion}
We have presented co-adaptive entropy modeling and entropy-coding-aware
perceptual bit allocation without transmitted model side information; the
dynamic program is exact per block under the model's rate estimates. Under a
strict per-block rate, entropy-coding-aware bit allocation lowers total
objective perceptual distortion by 15--24\,\% relative to the same solver
without it, and co-adaptation removes a further 3--4\,\% that a
backward-adaptive coder alone cannot realize in one pass; both gains hold
in- and out-of-domain. With a reservoir of one block or more, the two
increments shrink to 1--3\,\% and 0.3--0.5\,\%, so the co-adaptive gain is
realized under every tested reservoir policy. The Lagrangian solver stays
within about 0.5\,\% of the exact DP at about $1/20$ of its encode CPU time.
In the MUSHRA test at 80 kbps/channel with an unbounded reservoir, both
exact-DP conditions score 22--25 points above water-filling, and co-adaptive
DP scores 3.4 points above static DP. Natural extensions are rate smoothing
under a bounded reservoir and a faster exact solver.

\FloatBarrier
\Section{References}
\bibliographystyle{IEEEtran}
\bibliography{refs}

\end{document}